\documentclass[aps, prl, floatfix, twocolumn, amsmath, superscriptaddress, showpacs]{revtex4-1} 
\usepackage{graphicx}  
\usepackage{epstopdf}
\usepackage{amsfonts}

\newcommand{\rvec}{\textbf{r}}  
\newcommand{\rvecp}{\textbf{r}^{\prime}}  
\begin{document}

\title{Chimera states on a flat torus} 

\author{Mark J. Panaggio}
\email[email: ]{markpanaggio2014@u.northwestern.edu}
\affiliation{Department of Engineering Sciences and Applied Mathematics, Northwestern University, Evanston, Illinois 60208, USA}

\author{Daniel M. Abrams}
\affiliation{Department of Engineering Sciences and Applied Mathematics, Northwestern University, Evanston, Illinois 60208, USA}
\date{ \today}

\begin{abstract}
Discovered numerically by Kuramoto and Battogtokh in 2002, chimera states are spatiotemporal patterns in which regions of coherence and incoherence coexist.  These mathematical oddities were recently reproduced in a laboratory setting sparking a flurry of interest in their properties. Here we use asymptotic methods to derive the conditions under which two-dimensional chimeras, similar to those observed in the experiments, can appear in a periodic space.  We also use numerical integration to explore the dynamics of these chimeras and determine which are dynamically stable.
\end{abstract}
\maketitle

In nature, arrays of oscillators often synchronize and oscillate with a single frequency.  This phenomenon can be observed in diverse systems ranging from laser arrays~\cite{Wang1988} and Josephson junctions~\cite{Phillips1993, Watanabe1994, Wiesenfeld1998}, to populations of fireflies~\cite{Ermentrout1991} and heart cells~\cite{Michaels1987}.  While incoherence and synchronization are ubiquitous in these systems, other complex patterns are also possible.

The standard mathematical paradigm for modeling arrays of coupled oscillators is the Kuramoto model~\cite{Acebron2005}. Kuramoto approximated the complex Ginzburg-Landau equation and demonstrated that under weak coupling, amplitude changes can be ignored~\cite{Kuramoto2003}.  Thus, oscillators can be approximated as coupled only through their phase with dynamics governed by 
\begin{equation*}
	\dot{\phi_i}=\omega_i-K\sum\limits_{j=0}^N\sin \left(\phi_i-\phi_j+\alpha\right). 
\end{equation*}
where $\phi_i$ is the phase of oscillator $i$, $\omega_i$ is the natural frequency, $K$ is the coupling strength, and $\alpha$ is the coupling lag~\cite{Lakshmanan2010}.  For arrays with narrow unimodal natural frequency distributions and no lag ($\alpha=0$), a first order phase transition occurs.  Below a critical coupling strength, oscillators remain incoherent and above this threshold, they begin to synchronize~\cite{Strogatz2000,Ott2008}.

\begin{figure*}[bth!]
	\includegraphics[width=0.7\textwidth]{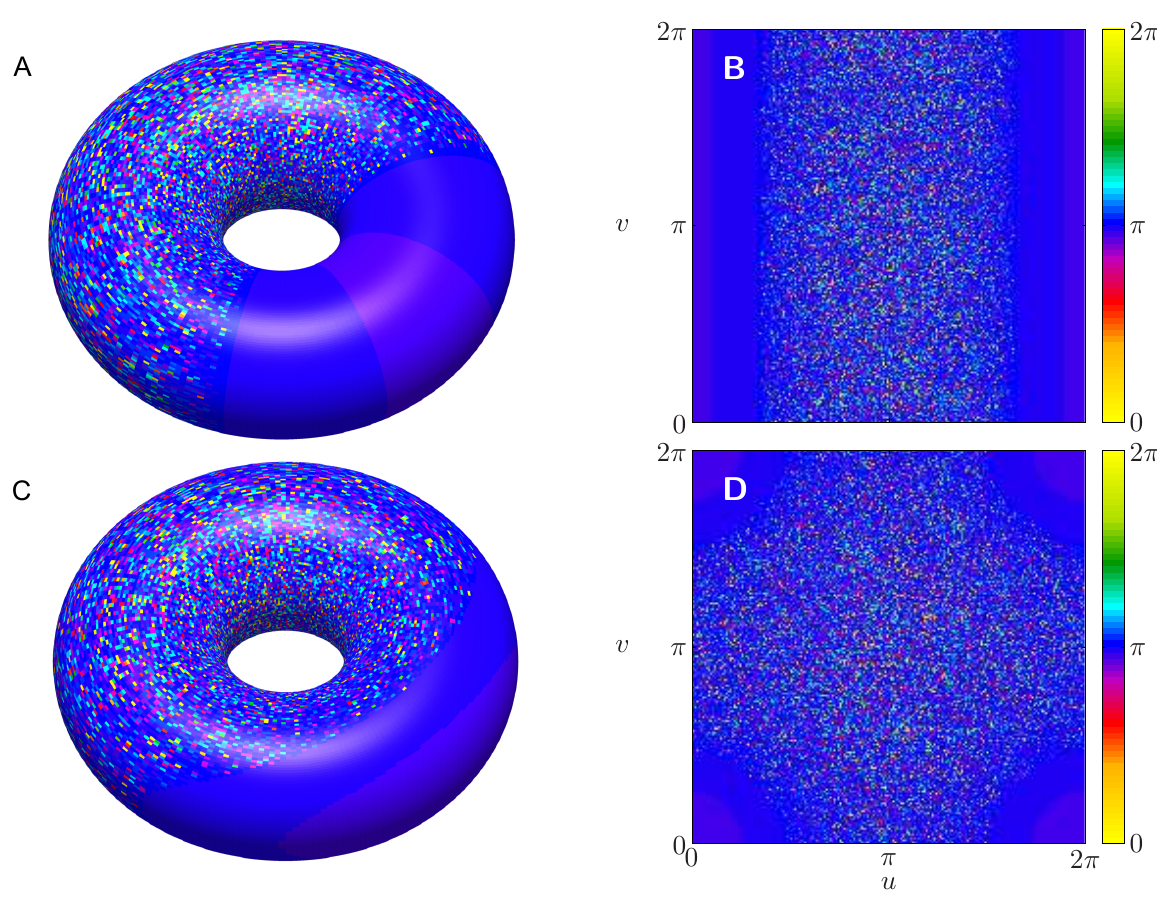}
	\caption{Stripe and symmetric spot chimeras.  Each panel contains a 200 by 200 grid of oscillators.  Each pixel corresponds to a single oscillator with color representing the phase.  Panels A and B display a stripe chimera state with $a_1=0.2322$, $a_2=0$, $\delta=0.1065$, while C and D display a symmetric chimera state with a coherent spot and with $a_1=0.1366$, $a_2=0.1366$, $\delta=0.1494$.  Numerical integration indicates that these states are stable.} 
	\label{fig:1d_sym_chimeras}
\end{figure*} 

In 2002, Kuramoto and Battogtokh observed that with nonzero lag and nonlocal coupling, surprisingly, regions of coherence and incoherence can coexist even for identical oscillators ($\omega_i=\omega\, \forall\, i$)~\cite{Kuramoto2002}. Abrams and Strogatz described this hybrid state as a ``chimera''~\cite{Abrams2004}. Since then, chimera states have been observed in various systems including: two groups of oscillators with no spatial extent~\cite{Abrams2008,Laing2009}, a ring of oscillators~\cite{Kuramoto2002,Abrams2004,Abrams2006}, an infinite plane~\cite{Shima2004, Martens2009}, and a periodic two-dimensional space~\cite{Omelchenko2012}.  These states display a variety of spatial patterns including stripes, spots (see Fig.~\ref{fig:1d_sym_chimeras}), and spirals.

Recently, two experiments observed chimeras in laboratory settings for the first time~\cite{Smart2012}. Tinsley, Nkomo and Showalter used photo-excitatory feedback to couple two populations of discrete chemical oscillators~\cite{Tinsley2012}.  Occasionally, one population synchronized while the other remained incoherent. This is consistent with the analytical results in ref.~\cite{Abrams2008}.   

Meanwhile, Hagerstrom et al.~used a computer with feedback from a camera to control the phase modulation induced by a spatial light modulator~\cite{Hagerstrom2012}. This created a physical realization of a two-dimensional iterated map with nonlocal coupling and periodic boundary conditions. They observed the formation of chimera states as one-dimensional stripes of incoherence.  Omel'chenko et al.~studied a similar system and produced both symmetric spot and stripe patterns in numerical experiments~\cite{Omelchenko2012}. These patterns have yet to be explained from an analytical perspective.

\textit{Analysis}---To determine conditions for the existence of these chimera states, we examine a two-dimensional array of oscillators in a space with periodic boundaries:
\begin{equation*}
\mathbb{T}^2 = S^1\times S^1=\{(u,v)| u\in [0,2\pi),v\in [0,2\pi)\}. 
\end{equation*}
This can be interpreted as the surface of a torus where the $u$- and $v$-coordinates correspond to the toroidal and poloidal angles (we disregard any effects of surface curvature).  

We consider a generalization of the traditional Kuramoto model to a continuous distribution of oscillators:
\begin{equation}
\label{kuramoto}
\frac{\partial\phi(\rvec,t)}{\partial t}=\omega-\int_{\mathbb{T}^2}G(\rvec,\rvecp )\sin (\phi(\rvec,t)-\phi(\rvecp,t)+\alpha)d\rvecp,
\end{equation}
where $G(\rvec,\rvecp)$ is a continuous coupling kernel. Following the approach of Kuramoto and Battogtokh~\cite{Kuramoto2002}, we shift into a rotating frame with angular frequency $\Omega$ (to be determined later) and define a complex order parameter, 
\begin{equation}
\label{orderparameter}
R(\rvec,t)e^{i\Theta(\rvec,t)}=\int_{\mathbb{T}^2}G(\rvec,\rvecp)e^{i\theta(\rvec',t)}d\rvecp,
\end{equation}
resulting in a new governing equation:
\begin{equation}
\label{governing_eq}
\frac{\partial\theta(\rvec,t)}{\partial t}=\Delta-R(\rvec,t)\sin (\theta(\rvec,t)-\Theta(\rvec,t)+\alpha),
\end{equation}
where $\theta=\phi-\Omega t$ and $\Delta=\omega-\Omega$. 

In regions where $R(\rvec)\geq|\Delta|$, stationary solutions to Eq.~\eqref{governing_eq} exist. Oscillators become phase-locked and rotate at a fixed angular frequency.  In regions where $R(\rvec)<|\Delta|$, stationary solutions are not possible. Instead, oscillators drift at a nonzero phase velocity and satisfy a stationary probability density.   Kuramoto and Battoogtokh observed that Eq.~\eqref{governing_eq} can be used to eliminate $e^{i\theta(\rvec,t)}$ from Eq.~\eqref{orderparameter}.  This yields a self-consistency equation for solutions to Eq.~\eqref{governing_eq}:
\begin{equation}
\label{self_c}
R(\rvec)e^{i\Theta(\rvec)}=e^{i\beta}\int_{\mathbb{T}^2}G(\rvec,\rvecp)h(\rvecp)e^{i\Theta(\rvec',t)}d\rvecp,
\end{equation}
where $h(\rvec)=\frac{\Delta -\sqrt{\Delta^2-R^2(\rvec)}}{R(\rvec)}$. 

This functional self-consistency equation is effectively infinite-dimensional and yields little insight.  To proceed, we define a simple kernel representing nonlocal coupling:
\begin{equation*}
G(\rvec,\rvecp)= \frac{1}{(2\pi)^2} \left[1+\kappa(\cos(u-u^\prime)+\cos(v-v^\prime))\right]. 
\end{equation*}
This is the leading order approximation to the two-dimensional von Mises distribution, the circular analogue of the Gaussian, and can also be interpreted as a perturbation off of all-to-all coupling for $\kappa\ll 1$. This choice allows us to remove explicit dependence on $u$ and $v$ from the integrals and express the order parameter as follows:
\begin{equation}
\label{self_c_2}
R(\rvec)e^{i\Theta(\rvec)}= c+d_1 \cos(u) +d_2 \cos(v),
\end{equation}
where $\langle f(\rvecp)\rangle = \frac{1}{(2\pi)^2}\int_{\mathbb{T}^2} f(\rvecp)d\rvecp$ and
\begin{subequations}\label{sc_alg}
\begin{align}
c&=e^{i\beta}\langle h(\rvecp)e^{i\Theta(\rvecp)}\rangle \label{sc_alg_c}\\
d_1&=\kappa e^{i\beta}\langle h(\rvecp)e^{i\Theta(\rvecp)}\cos(u^{\prime})\rangle \label{sc_alg_d1}\\
d_2&=\kappa e^{i\beta}\langle h(\rvecp)e^{i\Theta(\rvecp)}\cos(v^{\prime})\rangle. \label{sc_alg_d2}
\end{align}
\end{subequations}

Without loss of generality, we define $\Theta=0$ at the point $(u_0,v_0)=(\frac{\pi}{2},\frac{\pi}{2})$, resulting in real-valued $c$. We use Eq.~\eqref{self_c_2} to eliminate $R$ and $\Theta$ from system \eqref{sc_alg}, thus reducing infinite dimensional functional Eq.~\eqref{self_c} to a set of six algebraic equations (3 real and 3 complex) with six variables ($c$, $\textnormal{Re}(d_1)$, $\textnormal{Im}(d_1)$, $\textnormal{Re}(d_2)$, $\textnormal{Im}(d_2)$ and $\Delta$):
\begin{subequations}\label{sc_alg_v2}
\begin{align}
c&=e^{i\beta}\Bigg\langle \frac{\Delta -\sqrt{\Delta^2-\left|c+d_1 \cos(u^{\prime}) +d_2 \cos(v^{\prime})\right|^2}}{c^*+d_1^* \cos(u^{\prime}) +d_2^* \cos(v^{\prime})}\Bigg\rangle \label{eqc0}\\
d_1&=\kappa e^{i\beta}\Bigg\langle \frac{\Delta -\sqrt{\Delta^2-\left|c+d_1 \cos(u^{\prime}) +d_2 \cos(v^{\prime})\right|^2}}{c^*+d_1^* \cos(u^{\prime}) +d_2^* \cos(v^{\prime})}\cos(u^{\prime})\Bigg\rangle \label{eqc1}\\
d_2&=\kappa e^{i\beta}\Bigg\langle \frac{\Delta -\sqrt{\Delta^2-\left|c+d_1 \cos(u^{\prime}) +d_2 \cos(v^{\prime})\right|^2}}{c^*+d_1^* \cos(u^{\prime}) +d_2^* \cos(v^{\prime})}\cos(v^{\prime})\Bigg\rangle, \label{eqc2}
\end{align}
\end{subequations}
where $^*$ denotes complex conjugation. 

\begin{figure}[b!]
	\includegraphics[width=0.45\textwidth]{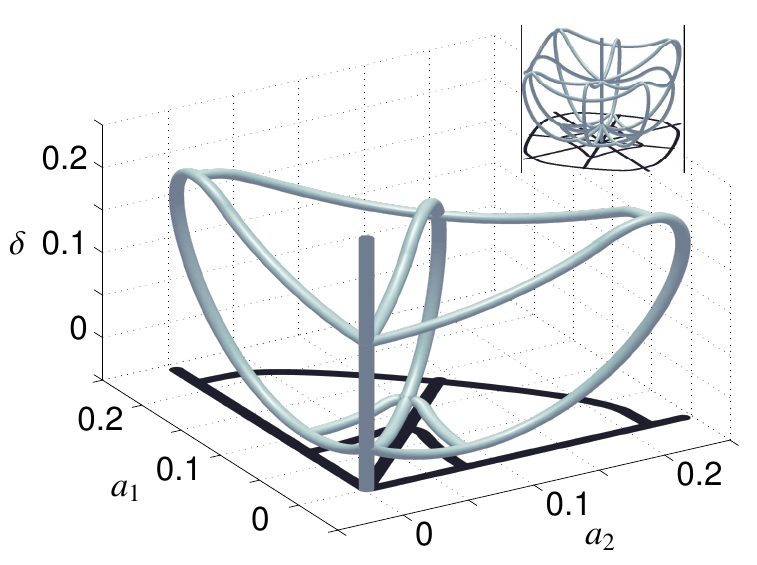}
	\caption{Solutions to system \ref{sc_asym_sys}.  The inset displays the full set of solutions. The main panel shows the quadrant containing only $a_1,a_2\geq0$.  Loops correspond to symmetric spot chimeras ($a_2=a_1$) and two types of equivalent stripe chimeras ($a_1=0$ and $a_2 =0$).  Branches ($a_1\neq a_2$, both nonzero) correspond to two types of asymmetric spot chimeras, one with $\delta>0$ and one with $\delta<0$.}
	\label{fig:limabean_eq}	 
\end{figure}

System \eqref{sc_alg_v2} can be solved asymptotically for $\kappa =\epsilon\ll 1$. Motivated by results from one dimension~\cite{Abrams2006}, we make an ansatz for the scalings
\begin{subequations}\label{scalings}
\begin{align}
\beta&=\beta_1\epsilon\\
c&\sim 1+c_1\epsilon+c_2\epsilon^2 \\
d_1&\sim (a_1+ib_1)\epsilon^2  \\
d_2&\sim(a_2+ib_2)\epsilon^2 \\
\Delta &\sim 1+\Delta_1\epsilon+\Delta_2\epsilon^2,
\end{align}
\end{subequations}
substitute these expressions into Eq.~\eqref{eqc0}, and retain terms to $\mathcal{O}(\sqrt{\epsilon})$:
\begin{equation*}
1+\mathcal{O}(\epsilon)=1+\sqrt{2}\sqrt{\Delta_1-c_1}\sqrt{\epsilon}+\mathcal{O}(\epsilon).
\end{equation*} 
For this to be satisfied, $\Delta_1=c_1$ is required. Expanding system \eqref{sc_alg_v2} to leading order we obtain
\begin{subequations} \label{sc_asym_sys}
\begin{align}
	c_1&=i\beta_1-\sqrt{2}\bigg\langle \sqrt{\delta-a_1 \cos(u^{\prime})-a_2 \cos(v^{\prime})}\bigg\rangle \label{eqc0_a}\\
	a_1+ib_1&=-\sqrt{2}\bigg\langle \cos(u^{\prime}) \sqrt{\delta-a_1 \cos(u^{\prime})-a_2 \cos(v^{\prime})}\bigg\rangle \label{eqc1_a}\\
	a_2+ib_2&=-\sqrt{2}\bigg\langle \cos(v^{\prime})\sqrt{\delta-a_1 \cos(u^{\prime})-a_2 \cos(v^{\prime})}\bigg\rangle \label{eqc2_a}
\end{align} 
\end{subequations}
where we have defined $\delta=\Delta_2-c_2$ for convenience.  

These three complex equations contain six variables: $c_1$, $a_1$, $b_1$, $a_2$, $b_2$ and $\delta$, and one  parameter: $\beta_1$. Practically, it is easiest to parameterize the system by $\delta$, solve for $a_1$ and $a_2$ using the real parts of implicit Eqs.~\eqref{eqc1_a} and \eqref{eqc2_a}, and then deduce the remaining values.  This yields a set of solutions corresponding to various types of chimera and drifting states.

\begin{figure}[t!]
	\includegraphics[clip=true,width=0.45\textwidth]{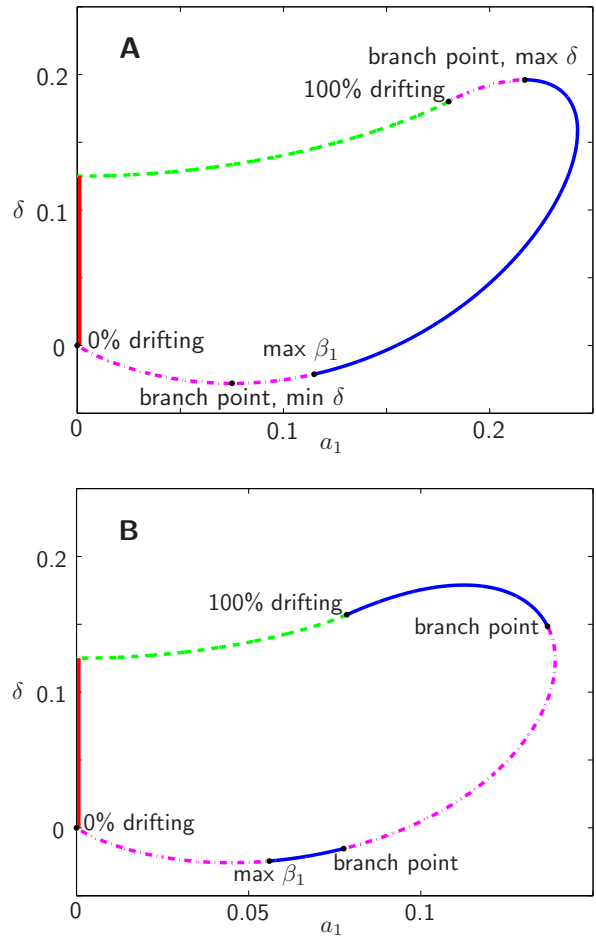}
	\caption{Cross sections of Fig.~\ref{fig:limabean_eq}.  The magenta (dash dot) curves represent unstable chimeras, the blue (solid) represent stable chimeras, the green (dashed) correspond to modulated drift states, and the red (solid, $a_1=0$) correspond to uniform drift states.  Panel A displays the stripe chimera loop ($a_2=0$). Panel B displays the symmetric spot loop ($a_1=a_2$).  
} 
	\label{fig:sc_cross_sections}
\end{figure} 

\textit{Results}---Fig.~\ref{fig:limabean_eq} shows solutions to system \eqref{sc_asym_sys}. It contains three closed loops connected by two branches.  The loops lie in the planes $a_1=0$, $a_2=0$, and $a_1=a_2$ whereas the branches have $a_1\neq a_2$ (both nonzero). Because Eq.~\eqref{self_c_2} is invariant under the transformations $(d_1,u)\rightarrow (-d_1,u-\pi)$, $(d_2,v)\rightarrow (-d_2,v-\pi)$, and $(d_1,u)\rightarrow (d_2,v)$, the quadrant with nonnegative $a_1$ and $a_2$ contains all of the distinct solutions to system \eqref{sc_asym_sys}, and the loops satisfying $a_1=0$ and $a_2=0$ are essentially the same.

\begin{figure*}[t!]
			\includegraphics[clip=true,width=0.7\textwidth]{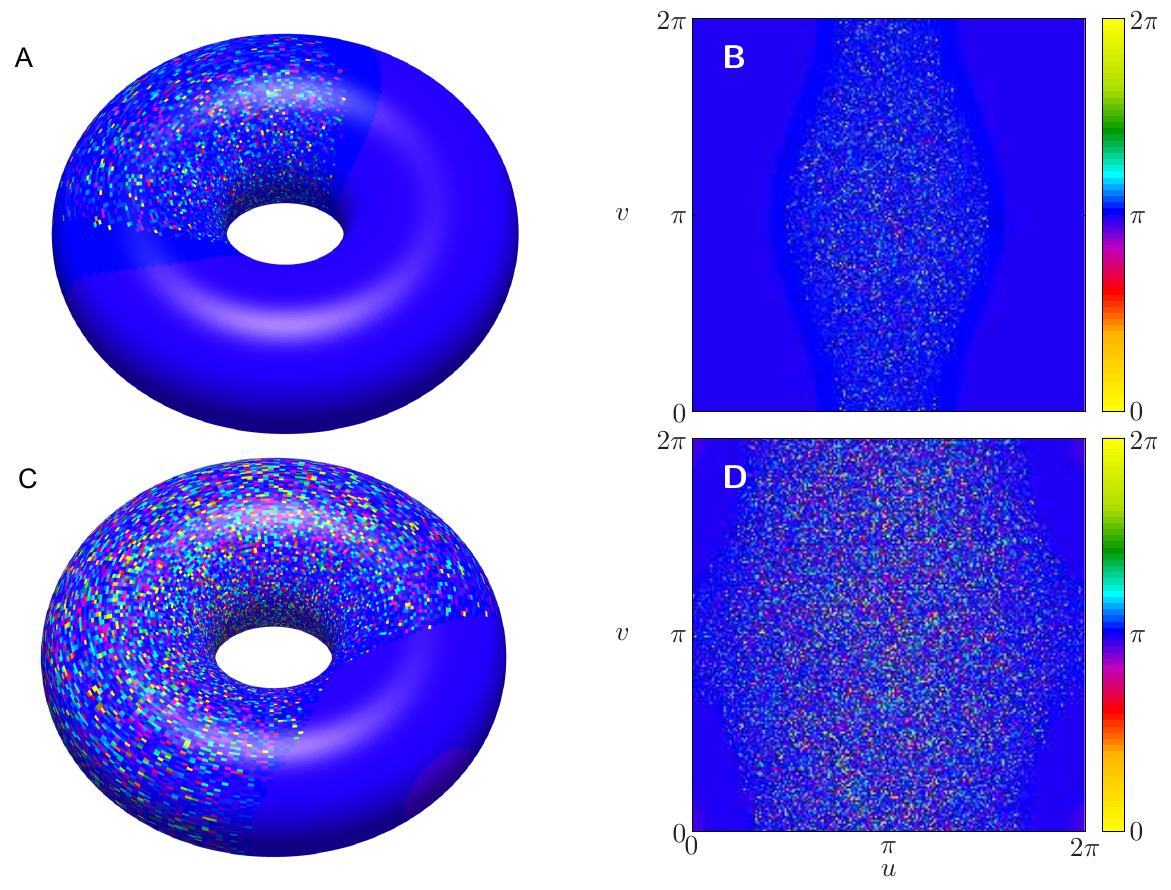}
	\caption{Asymmetric spot chimeras. Panels A and B display an asymmetric chimera from the lower branch with $a_1=0.07635$, $a_2=0.02929$, $\delta=-0.02608$, while C and D display an asymmetric chimera state from the upper branch with $a_1=0.2089$, $a_2=0.04844$, $\delta=0.1769$. Numerical integration reveals that these states are unstable.} 
	\label{fig:asym_chimeras}
\end{figure*}

The chimera states these solutions describe have locked and drifting regions separated by a contour where $R=\left|\Delta\right|$. Expressing this boundary in terms of the variables from Eq.~\eqref{scalings} yields \begin{equation}
\label{boundary_eq}
\delta = a_1\cos(u)+a_2\cos(v).
\end{equation} 
This boundary takes on various forms depending on the values of $a_1$, $a_2$ and $\delta$.

When $a_1=0$ or $a_2=0$, system \eqref{sc_alg_v2} reduces to the equations analysed in ref.~\cite{Abrams2006}. The solutions along this branch represent one-dimensional stripe chimeras---the order parameter varies in only one spatial dimension. One such chimera is displayed in Figs.~\ref{fig:1d_sym_chimeras}(A) and \ref{fig:1d_sym_chimeras}(B).  Fig.~\ref{fig:sc_cross_sections}(A) describes the various solutions that are found along this branch and indicates their stability. These chimeras are qualitatively similar to those observed in experiments~\cite{Hagerstrom2012}. 

When $a_1=a_2$, the boundary is symmetric about the line $u=v$ and resembles a circle or a square. These chimeras consist of an incoherent spot surrounded by a coherent region or a coherent spot surrounded by an incoherent region, similar to those studied in ref.~\cite{Omelchenko2012}.  An example of such a state can be found in Figs.~\ref{fig:1d_sym_chimeras}(C) and \ref{fig:1d_sym_chimeras}(D). Solutions along this loop are described in Fig.~\ref{fig:sc_cross_sections}(B).

When $a_1\neq a_2$, the boundary resembles an ellipse or a rhombus, both of which are asymmetric (they are not invariant under reflections about the line $u=v$). There are two such branches, one with $\delta>0$ and one with $\delta<0$.  These solutions also have regions of incoherence surrounding or surrounded by coherence. Examples of these previously undiscovered chimeras are found in Fig.~\ref{fig:asym_chimeras}. 

\begin{figure}[ht]
	\includegraphics[width=0.45\textwidth]{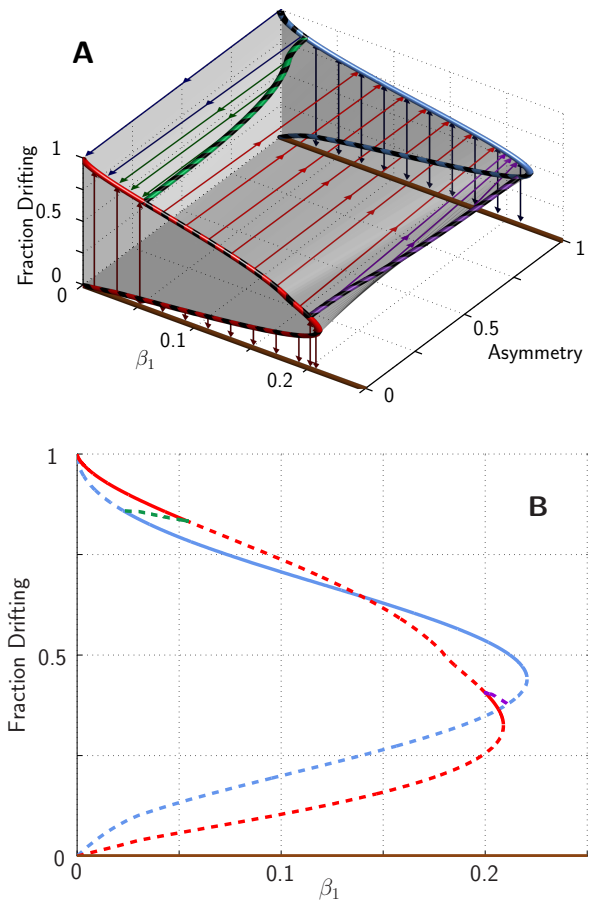}
	\caption{Dynamics of the fraction drifting and asymmetry.  Solid lines indicate stable states while dashed lines indicate unstable states. Red curve: symmetric spot chimeras, blue curve: stripe chimeras, green curve: asymmetric spot chimeras ($\delta>0$), purple curve: asymmetric spot chimeras ($\delta<0$).  Unstable states evolve along planes of fixed $\beta_1$ to nearby stable states (with a different fraction drifting and/or asymmetry, see text for definition) as indicated by the arrows in panel A.  Panel B contains a two-dimensional projection of panel A showing the fraction drifting as a function of $\beta_1$ for the various chimeras.} 
	\label{fig:beta_drift}
\end{figure}

We determined the stability of predicted chimera states by numerically integrating Eq.~\eqref{kuramoto} with initial conditions determined by our theory. Fig.~\ref{fig:beta_drift} plots the fraction of oscillators in the drifting region and the asymmetry of that region (defined as min$\left(\left|a_1/a_2\right|,\left|a_2/a_1\right|\right)$) as a function of $\beta_1$. The stripe chimera loop contains both stable and unstable domains, and is nearly identical to the solution curve described in ref.~\cite{Abrams2006} (see Fig.~\ref{fig:sc_cross_sections}(A)).  The symmetric spot loop also contains both stable and unstable domains  (see Fig.~\ref{fig:sc_cross_sections}(B)) and is similar in shape to the stripe chimera loop, but it has differing stability regions. Both asymmetric branches are unstable, and nearby states evolve within planes of fixed $\beta_1$ to solutions along the stripe or symmetric spot loops.         

\textit{Conclusions}---This work reveals a new type of chimera state in a two-dimensional periodic space.  These asymmetric chimeras are unstable and appear in simulations only as transients.  Nonetheless, understanding where these asymmetric chimeras appear is essential to understanding where stripe and spot chimeras are stable. Stable chimeras are created via continuous bifurcation off of modulated drift states, while unstable chimeras appear via continuous bifurcation off of the fully synchronized state.  As $\beta_1$ increases, symmetric and stripe loops intersect with asymmetric branches resulting in unexpected changes in stability.  Eventually, a stable chimera collides with an unstable chimera causing both to be destroyed in a saddle-node bifurcation.  

Our analysis reveals the complex regions of parameter space in which two-dimensional chimeras reside.  The intricacy of bifurcation diagrams \ref{fig:sc_cross_sections} and \ref{fig:beta_drift} elucidate why it is difficult to reproduce chimera states in numerical simulations and experiments.   Stable chimeras only exist for narrow ranges of the system parameters.  Consequently, small changes in the lag parameter $\alpha$ (or equivalently $\beta$), can cause stable chimeras to vanish or new chimeras to appear. 

These findings also highlight the impact of topology on equilibrium states.  On a torus, single spirals are excluded by the periodic boundaries---they are topologically impossible.  However, on an infinite plane, finite-sized non-spiral chimeras have not yet been observed and may not exist.  Networks of oscillators are often of interest, and do not necessarily reproduce the topological properties of any simple metric space.  We hypothesize that, on arbitrary networks of oscillators with more complex structure \cite{Restrepo2005}, previously unobserved chimeras are possible.  Investigating these chimeras may shed light on breakdowns of synchrony observed in nature.

\end{document}